\documentstyle[11pt,newpasp,twoside]{article}
\def\msun{M$_{\odot} \ $}
\def\I{{\'\i}}

\def\edcomment#1{\iffalse\marginpar{\raggedright\sl#1\/}\else\relax\fi}
\reversemarginpar

\begin{document}
\title{Photometric Constraints on Chemical Evolution Models of Irregular Galaxies}
 \author{Leticia Carigi}
\affil{ Instituto de Astronom\I a, UNAM, Apdo. Postal 70-264, CP 04510, M\'exico DF, Mexico}
\author{Gustavo Bruzual}
\affil{Centro de Investigaciones de Astronom\I a, Apdo. Postal 264, M\'erida 5101-A, Venezuela}

\begin{abstract}
Photometric properties of stellar populations that evolve according to chemical evolution models 
developed by Carigi, Col\I n, \& Peimbert (1999) and Carigi \& Peimbert (2001) are explored. Models 
explain the oxygen abundance and gas mass fraction of irregular galaxies NGC 1560, I Zw 18, NGC 2366, 
and a typical irregular galaxy. The photometric predictions help to narrow down the range of possible
 chemical evolution models. Observed colors of I Zw 18 imply an age of $10^8 - 10^9$ yr  for its 
dominant population, while $10^9 - 10^{10}$ yr is an older age preferred by the rest of the 
galaxy sample. 
Observed colors are in general, within errors, close to predicted. There is a tendency for the observed
$(B-V)$ color to imply higher metallicities at a fixed age than $(U-B)$ (equivalently, $(U-B)$
 implies younger
ages than $(B-V)$).  

\end{abstract}

\section{Introduction}
Irregular galaxies are metal poor systems whose oxygen deficiency can  be explained by
an IMF with a   
 larger fraction of
 low-mass stars
relative to the solar vicinity IMF,  but similar to that found in globular clusters
(Carigi, Col\I n, \& Peimbert 1999, CCP; Carigi \& Peimbert 2001, CP).
In both papers, the chemical evolution of 
three irregular galaxies and an average galaxy called ``typical irregular galaxy"
are discussed by models for three ages: $10^8$, $10^9$, and $10^{10}$ yr
and different factors $r$ that indicate the relative excess of sub-stellar objects ($m<0.1$ \msun).
The lack of abundance determinations for other elements besides O, or an
estimation of gas mass fraction, prevented the authors to discriminate which of these
models ($r$, age) is most appropriate.

This work explores whether existing photometric data for these galaxies and predictions from population
synthesis models (Carigi \& Bruzual 2000) can indicate the chemical
history of these irregular galaxies, under the CCP and CP chemical evolution models.

\section{Results and Discussion}

Carigi \& Bruzual (2000) adopt IMF (or factor $r$) and $Z(t)$ from closed-box models
with continuous or bursting star formation rates.

($U-B$) and ($B-V$) colors evolved according to different spectral evolution models 
and color data
suggest that both, {\bf NGC 1560} and {\bf NGC 2366} have been forming stars for the last 10 Gyr.
The case {\bf I Zw 18} is different, its observed colors are consistent with an age between $10^8$ and $10^9$ yr,
certainly lower than for NGC 1560 and NGC 2366.

For the {\bf Typical Irregular Galaxy} 
 the $(U-B)$ color is bluer for a given $(B-V)$ than expected
 from continuous star formation models. This may also indicate that star formation in these systems
 occurs in bursts. After each burst of star formation, a galaxy becomes bluer (especially in $(U-B)$)
 due to the increase in the number of massive main sequence. CCP models with bursting SFR can explain
the bluest galaxies in the sample, as well as the O abundance of this system. 

We compares observed colors of several {\bf Local Group dwarf irregular galaxies}, with
models for NGC 1560, the best known galaxy in our sample.
Again, observed colors are consistent with the predictions
from chemical evolution models $10^9$ and $10^{10}$ yr old, 
although some galaxies show $(B-V)$ colors redder than expected.

\section {Conclusions}

In this work we have explored the color behavior of stellar populations that evolve according to 
chemical evolution models. Our main conclusions are:

$\diamond$ At a given age, the color predicted for a typical irregular galaxy is independent of the fraction 
of sub-stellar objects. Therefore, color data helps to discriminate model age but not 
$r$.

$\diamond$ Based on available photometric data, models with fast chemical enrichment (age less than $10^9$ yr) 
are discarded for most irregular galaxies. These populations are too young to reach the observed 
colors. The exception is I Zw 18, whose favored age is between $10^8$ and $10^9$ yr.

$\diamond$ Although the age of the dominant stellar population in these galaxies is unknown, photometric models
can be used to arrange these systems on an age sequence based on observed colors. We have found a
tendency of unknown origin for the $(B-V)$ color to be matched by stellar populations of higher $Z$
than the $(U-B)$ color, at constant age. Equivalently, $(U-B)$ implies younger ages, at a given
metallicity, than $(B-V)$.

$\diamond$ At a given age, color and abundance ratios are almost independent of the fraction of sub-stellar 
objects.

\end{document}